\documentclass[useAMS,usenatbib]{mn2e}

\usepackage[dvipdfm]{graphicx,color}
\usepackage[normalem]{ulem}
\usepackage{color}
\newcommand{\vo}{{V}_{0}}

\newcommand{\vio}{({V}-{I}_{c})_{0}}
\newcommand{\mv}{\rm{M}_{V}}
\newcommand{\mm}{(m-M)_{0}}

\newcommand{\lsolar}{\rm{L_{\odot}}}
\newcommand{\feh}{\rm{[Fe/H]}}
\newcommand{\afe}{\rm{[\alpha/Fe]}}

\newcommand{\rc}{\rm{R}_{c}}
\newcommand{\rt}{\rm{R}_{t}}
\newcommand{\rh}{\rm{R}_{h}}

\title[Population gradient in Sextans dSph]{Population gradient in Sextans dSph:\\ 
Comprehensive mapping of a dwarf galaxy by Suprime-Cam\thanks{Based on data collected at Subaru Telescope, which is operated by the National Astronomical Observatory of Japan}}
\author[S. Okamoto et al.]
	{S. Okamoto$^{1}$\thanks{E-mail: okamoto@shao.ac.cn}, 
	N. Arimoto$^{2,3,4}$, 
	E. Tolstoy$^{5}$, 
	P. Jablonka$^{6,7}$, 
	M. J. Irwin$^{8}$,
\newauthor
	Y. Komiyama$^{4}$, 
	Y. Yamada$^{4}$,
	and M. Onodera$^{2}$\\
	$^{1}$Shanghai Astronomical Observatory, Chinese Academy of Sciences, 80 Nandan Rd. Shanghai, 200030 China\\
	$^{2}$Subaru Telescope, 650 North Aohoku Place, Hilo, HI 96720, USA\\
	$^{3}$The Graduate University for Advanced Studies, Osawa 2-21-1, Mitaka, Tokyo 181-8588, Japan\\
	$^{4}$National Astronomical Observatory of Japan, Osawa 2-21-1, Mitaka, Tokyo 181-8588, Japan\\	
	$^{5}$Kapteyn Institute, University of Groningen, Postbus 800, 9700AV Groningen, the Netherlands\\
	$^{6}$Laboratoire d'astrophysique, Ecole Polytechnique F\'ed\'erale de Lausanne (EPFL), Observatoire de Sauverny, CH-1290 Versoix, \\Switzerland\\
	$^{7}$GEPI, Observatoire de Paris, CNRS, Universit\'e de Paris Diderot, F-92195 Meudon, Cedex, France\\
	$^{8}$Institute of Astronomy, University of Cambridge, Madingley Road, Cambridge CB3 0HA, U.K.}
	
\begin{document}

%\date{in original form 2007 June 15}

%\pagerange{\pageref{firstpage}--\pageref{lastpage}} \pubyear{2007}
\maketitle

\label{firstpage}

\begin{abstract}
We present the deep and wide $V$ and $I_c$ photometry of the Sextans dwarf spheroidal galaxy (dSph) taken by Suprime-Cam imager on the Subaru Telescope, which extends out to the tidal radius.  The colour-magnitude diagram (CMD) reaches two magnitudes below the main sequence (MS) turn-off, showing a steep red giant branch, blue and red horizontal branch (HB), sub-giant branch (SGB), MS, and blue stragglers (BS).  We construct the radial profile of each evolutionary phase and demonstrate that blue HB stars are more spatially extended, while red HB stars are more centrally concentrated than the other components.  The colour distribution of SGB stars also varies with the galactocentric distance; the inner SGB stars shift bluer than those in the outskirt.  
The radial differences in the CMD morphology indicate the existence of the age gradient.  The relatively younger stars ($\sim10$ Gyr) are more centrally concentrated than the older ones ($\sim13$ Gyr).  The spatial contour maps of stars in different age bins also show that the younger population has higher concentration and higher ellipticity than the older one.  We also detect the centrally concentrated bright BS stars, the number of which is consistent with the idea that a part of these stars belongs to the remnant of a disrupted star cluster discovered in the previous spectroscopic studies. 

\end{abstract}


\section{Introduction}
Dwarf galaxies are the most simple and numerous galaxies in the Universe and their stellar populations provide clues about the formation and evolution of galaxies in various environments.  Within the currently favoured cosmological Cold Dark Matter scenario, stellar halos of large galaxies were built up through mergers of low-mass systems, such as the dwarf spheroidal (dSph) galaxies around the Milky Way.

Since 2005, analyses of resolvable stellar distribution based on large photometric survey programs have led to discoveries of numerous ultra faint galaxies (UFDs) \citep[e.g.][]{2005ApJ...626L..85W, 2006ApJ...647L.111B}.  The deep colour-magnitude diagrams (CMDs) show that they have quite an old and metal-poor population \citep[e.g.][]{2008A&A...487..103O, 2012ApJ...744...96O, 2012ApJ...753L..21B}.  On the other hand, radial population gradients and distinct dynamical components are found in many brighter dwarf satellites, such as Fornax and Sculptor dSph \citep[e.g.][]{2004ApJ...617L.119T, 2006A&A...459..423B, 2008ApJ...681L..13B}. The differences in stellar populations, chemical compositions and the structural properties between the classical dSphs and UFDs give hints to understand the star formation in low mass systems in the early Universe as well as the formation of the Milky Way and its satellites. 

The Sextans dSph was discovered from UK Schmidt Telescope (UKST) sky survey \citep{1990MNRAS.244P..16I}.  It has the lowest central surface brightness ($\Sigma_0=18.2\pm0.5\,\rm{mag}\,\rm{arcmin}^{-2}$) of the Galactic dSph, excluding the most recently discovered ones, and has a high mass-to-light ratio ($M/L_{V} \sim 40 M_{\odot}/L_{\odot ,V}$) \citep{1991AJ....101..892M, 1998ARA&A..36..435M}. It lies at a distance of $94 \pm 8$ kpc and its tidal radius estimates vary between $\sim 80$ and $\sim 160 \arcmin$ \citep{1995MNRAS.277.1354I, 2016MNRAS.460...30R}. Due to the large extent in the sky, it is difficult to study the global stellar population by most current cameras and spectrographs on large aperture telescopes.  The observational studies covering the central region showed some evidence of the presence of multiple stellar populations \citep[e.g.][]{2001MNRAS.327L..15B, 2001AJ....122.3092H, 2004MNRAS.354L..66K}.  \cite{2003AJ....126.2840L} (hereafter L03) presented a study of the wide-field BVI photometry reaching to the main-sequence (MS).  They covered the central $42\arcmin \times 28\arcmin$ region using CFH12K camera on the Canada-France-Hawaii Telescope and reported the different spatial distributions of red and blue horizontal branch stars (HB); the red HB stars are centrally concentrated and blue HB stars are more extended.  Subsequently, a synthetic model was adapted to their photometric catalogue to reveal the star formation history (SFH) and chemical enrichment history as a function of a distance from the centre \citep{2009ApJ...703..692L}.  It showed that, within the half-light radius of Sextans, star formation was more effective and lived longer in the central regions than the outer regions. More recently, \cite{2016MNRAS.460...30R} investigated the area within $4\times\rh$ of Sextans using DECam on CTIO 4m Blanco telescope, and found substructures at around $\rt$. 

The medium-resolution spectroscopic studies (MSR), based on both CaII triplet and the synthetic methods, revealed that the radial gradient of the metallicity is very small in the Sextans dSph, but the metallicity dispersion decreases with the distance from the centre \citep{2011MNRAS.411.1013B, 2011ApJ...727...78K}.  The inner region ($\rm{r}<0.8^{\circ}$) shows the whole range of $\feh$ from $-3.2$ to $-1.4$, with an average $\feh = -1.9$, while at larger radii only stars of more poor than $\feh \sim -2.2$ are present\citep{2011MNRAS.411.1013B}.  High resolution spectroscopic studies of a number of red giant branch (RGB) stars confirmed the range of metallicities estimated by those MSR studies \citep{2001ApJ...548..592S, 2009A&A...502..569A, 2010A&A...524A..58T, 2011PASJ...63S.523H}.  These photometric and spectroscopic studies indicate that the star formation period is more extended in the central regions than in the outer region, at least within the observed area.  

\cite{2004MNRAS.354L..66K} found a cold substructure in the central $5\arcmin$ of Sextans, from seven stars with radial velocity measurements.  The subsequent study with a larger spectroscopic sample confirmed this central substructure, the total luminosity and metallicity of which were estimated as $2.2\times 10^{4}\lsolar$ and $\feh=-2.6\pm0.15$, respectively \citep{2011MNRAS.411.1013B}.  \cite{2006ApJ...642L..41W} also found another kinematically cold substructure of $\sim10^{4}\lsolar$ luminosity at around the core radius, while they did not confirm the central one. 

In this paper, we present the deep and wide field photometry of the Sextans dSph covering entire region within $2\times\rc$ (or $1.5\times\rh$) and sampling fields to beyond the tidal radius.  Using Subaru/Suprime-Cam, we reveal the spatial differences of stellar populations from the innermost to the outskirt region.  The observation and data analysis procedures are described in Section 2.  In Section 3, we present the resulting CMDs, and derive the distance and the structural properties.  The spatial difference of the stellar population and the blue straggler (BS) stars are discussed in Section 4 and 5.  Finally we summarize our conclusion in Section 6.

\section{Observation and Data Reduction}

\subsection{Observation}

\begin{figure}
 \includegraphics[width=240pt,clip]{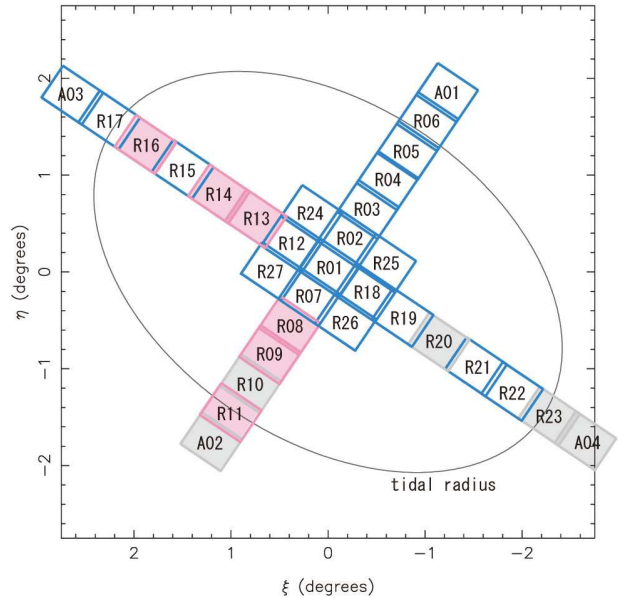}
 \caption{The target fields of Sextans dSph.  The box represents the field of view of the Suprime-Cam ($34\arcsec \times 27\arcsec$).  The blue and red boxes indicate our observed 26 fields which are good ($<$ 1.2\arcsec) and bad ($>$ 1.2\arcsec) seeing, respectively.  The gray boxes show the unobserved field.  The tidal radius from \citep{1995MNRAS.277.1354I} have been overlaid as a solid line. }
  \label{fig:1}
\end{figure}

We observed 26 fields in and around the Sextans dSph by using the prime focus imager Suprime-Cam \citep{2002PASJ...54..833M} on the Subaru Telescope during nights of December 30, 2005 to January 2, 2006 (PI. N. Arimoto), with the seeing ranging from 0.6\arcsec to 1.5\arcsec.  The Suprime-Cam camera provides a field-of-view of $34\arcmin\times27\arcmin$ with a pixel scale of 0.202\arcsec.  To avoid the saturation of bright stars, we took long and short exposure images with Johnson $V$ ($10\times50\,\rm{sec}$ and $3\times10\,\rm{sec}$) and Cousins $I$ ($10 \times110\,\rm{sec}$ and $3 \times30\,\rm{sec}$) filters. The details of the observations are summarized in Table~\ref{tbl:1}.

\begin{table}
\caption{Information about the observation and target fields}
\label{tbl:1}
\begin{tabular}{@{}lcccccc}
\hline\hline
	Field & Filter & \multicolumn{2}{c}{Short} & \multicolumn{2}{c}{Long} \\
\cline{3-4}
\cline{5-6}
		&	& time(sec) & FWHM & time(sec) & FWHM\\
\hline
R01 &  $I_c$ & 30$\times$3  & 0\farcs95 & 235$\times$10 & 0\farcs88 \\
    & $V$ & 10$\times$13 & 1\farcs06 & 50$\times$10  & 1\farcs02 \\
R02 &  $I_c$ & 30$\times$3  & 0\farcs98 & 235$\times$5  & 1\farcs06 \\
    & $V$ & 10$\times$13 & 1\farcs00 & 50$\times$5   & 0\farcs98 \\
R03 &  $I_c$ & 30$\times$3  & 0\farcs70 & 235$\times$5  & 0\farcs68 \\
    & $V$ & 10$\times$3  & 0\farcs99 & 50$\times$5   & 1\farcs02 \\
R04 &  $I_c$ & 30$\times$3  & 0\farcs70 & 235$\times$5  & 0\farcs66 \\
    & $V$ & 10$\times$3  & 0\farcs94 & 50$\times$5   & 0\farcs90 \\
R05 &  $I_c$ & 30$\times$3  & 0\farcs66 & 235$\times$5  & 0\farcs70 \\
    & $V$ & 10$\times$3  & 0\farcs92 & 50$\times$5   & 0\farcs77 \\
R06 &  $I_c$ & 30$\times$3  & 0\farcs64 & 235$\times$5  & 0\farcs94 \\
    & $V$ & 10$\times$3  & 0\farcs72 & 50$\times$5   & 0\farcs74 \\
R07 &  $I_c$ & 30$\times$3  & 0\farcs80 & 235$\times$5  & 0\farcs82 \\
    & $V$ & 10$\times$3  & 0\farcs78 & 50$\times$5   & 0\farcs77 \\  
R08 &  $I_c$ & 30$\times$3  & 0\farcs84 & 235$\times$5  & 0\farcs94 \\
    & $V$ & 10$\times$3  & 1\farcs20 & 50$\times$5   & 1\farcs34 \\ 
R09 &  $I_c$ & 30$\times$3  & 1\farcs14 & 235$\times$5  & 0\farcs80 \\
    & $V$ & 10$\times$3  & 1\farcs26 & 50$\times$5   & 1\farcs36 \\
R11 &  $I_c$ & 30$\times$3  & 0\farcs76 & 235$\times$5  & 0\farcs78 \\
    & $V$ & 10$\times$3  & 1\farcs34 & 50$\times$5   & 1\farcs38 \\
R12 &  $I_c$ & 30$\times$3  & 0\farcs92 & 235$\times$5  & 0\farcs80 \\
    & $V$ & 10$\times$3  & 1\farcs00 & 50$\times$5   & 0\farcs94 \\
R13 &  $I_c$ & 30$\times$3  & 0\farcs62 & 235$\times$5  & 0\farcs78 \\
    & $V$ & 10$\times$3  & 1\farcs14 & 50$\times$5   & 1\farcs34 \\
R14 &  $I_c$ & 30$\times$3  & 0\farcs70 & 235$\times$5  & 0\farcs78 \\
    & $V$ & 10$\times$3  & 1\farcs28 & 50$\times$5   & 1\farcs30 \\
R15 &  $I_c$ & 30$\times$3  & 0\farcs80 & 235$\times$5  & 0\farcs82 \\
    & $V$ & 10$\times$3  & 0\farcs80 & 50$\times$5   & 0\farcs86 \\
R16 &  $I_c$ & 30$\times$3  & 0\farcs82 & 235$\times$5  & 0\farcs82 \\
    & $V$ & 10$\times$3  & 1\farcs12 & 50$\times$5   & 1\farcs26 \\
R17 &  $I_c$ & 30$\times$3  & 0\farcs63 & 235$\times$5  & 0\farcs70 \\
    & $V$ & 10$\times$3  & 0\farcs76 & 50$\times$5   & 0\farcs82 \\
R18 &  $I_c$ & 30$\times$3  & 0\farcs86 & 235$\times$5  & 0\farcs84 \\
    & $V$ & 10$\times$3  & 1\farcs04 & 50$\times$5   & 0\farcs86 \\ 
R19 &  $I_c$ & 30$\times$6  & 0\farcs60 & 235$\times$5  & 0\farcs92 \\
    & $V$ & 10$\times$3  & 0\farcs70 & 50$\times$5   & 0\farcs72 \\
R21 &  $I_c$ & 30$\times$3  & 0\farcs96 & 235$\times$5  & 0\farcs90 \\
    & $V$ & 10$\times$3  & 0\farcs84 & 250$\times$3  & 0\farcs82 \\
R22 &  $I_c$ & 30$\times$3  & 0\farcs66 & 235$\times$5  & 0\farcs70 \\
    & $V$ & 10$\times$3  & 0\farcs76 & 50$\times$5   & 0\farcs90 \\
R24 &  $I_c$ & 30$\times$3  & 0\farcs90 & 235$\times$5  & 0\farcs86 \\
    & $V$ & 10$\times$3  & 0\farcs94 & 50$\times$5   & 0\farcs84 \\
R25 &  $I_c$ & 30$\times$3  & 1\farcs22 & 235$\times$5  & 1\farcs04 \\
    & $V$ & 10$\times$3  & 1\farcs08 & 50$\times$5   & 1\farcs02 \\ 
R26 &  $I_c$ & 30$\times$3  & 0\farcs78 & 235$\times$5  & 0\farcs86 \\
    & $V$ & 10$\times$3  & 0\farcs88 & 50$\times$5   & 0\farcs94 \\
R27 &  $I_c$ & 30$\times$3  & 0\farcs94 & 235$\times$5  & 0\farcs94 \\
    & $V$ & 10$\times$3  & 1\farcs06 & 50$\times$5   & 1\farcs08 \\
A01 &  $I_c$ & 30$\times$3  & 0\farcs68 & 235$\times$5  & 1\farcs04 \\
    & $V$ & 10$\times$3  & 0\farcs74 & 50$\times$5   & 0\farcs70 \\
A03 &  $I_c$ & 30$\times$3  & 0\farcs64 & 235$\times$5  & 0\farcs70 \\
    & $V$ & 10$\times$3  & 0\farcs88 & 50$\times$5   & 0\farcs92 \\

\hline\hline
\end{tabular}
\end{table}

In Figure \ref{fig:1}, the target fields (R01-R27) and the control fields (A01-A04) of the original observation plan are shown.  We chose nine fields at the galaxy centre and 18 fields along the direction of major and minor axis.  The blue and red boxes represent the observed 26 fields with good ($\rm{FWHM}< 1.25\arcsec$; blue) and bad ($\rm{FWHM}> 1.25\arcsec$; red) seeing conditions, respectively, and the gray boxes show the unobserved field.  In this paper, we use the regions which have $\rm{FWHM} < 1.25\arcsec$ in both $V$- and $I_c$-band images.  

\begin{figure}
 \includegraphics[width=170pt, angle=-90,clip]{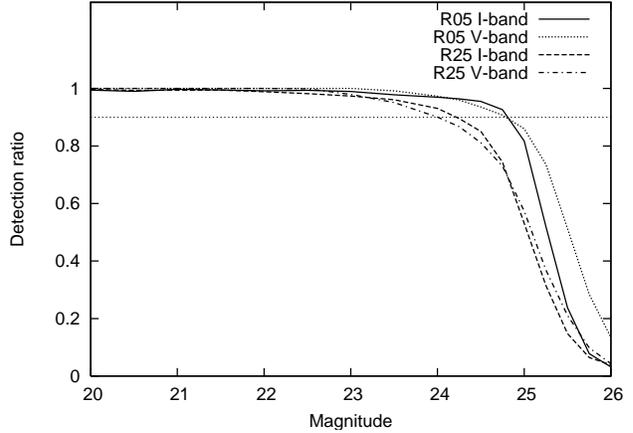}
 \caption{The completeness of detected stellar objects.  The solid and dotted lines are the detection ratio of I- and V-band images in the R05 field, respectively, which represent the case of good seeing condition ($\sim 0.7$\arcsec).  The dashed and dash-dotted lines indicate those in the R25 field which is the worst case used in the detailed analysis.  The 90\% complete-level is shown as a horizontal dotted line. }
  \label{fig:2}
\end{figure}

\subsection{Data reduction}
Data were processed using a pipeline software SDFRED dedicated to the Suprime-Cam \citep{2002AJ....123...66Y, 2004ApJ...611..660O}. Each image was processed and calibrated in the same manner as for Suprime-Cam images in \cite{2008A&A...487..103O}. For these processed images, the DAOPHOT in IRAF package was used to obtain the point-spread-function (PSF) photometry of the resolved stars \citep{1987PASP...99..191S}.  To separate the point sources from the extended ones and noise-like objects, we selected the sources having the DAOPHOT parameter $\chi^2$ and SHARP within 3$\sigma$ of the clipped-mean values of artificial stars at the same magnitude in the artificial-star test.  The positions of detected stellar objects in each processed image were cross-correlated (within 1\arcsec) to make a composite catalogue of long and short exposure images.   

\begin{figure*}
 \includegraphics[width=450pt,clip]{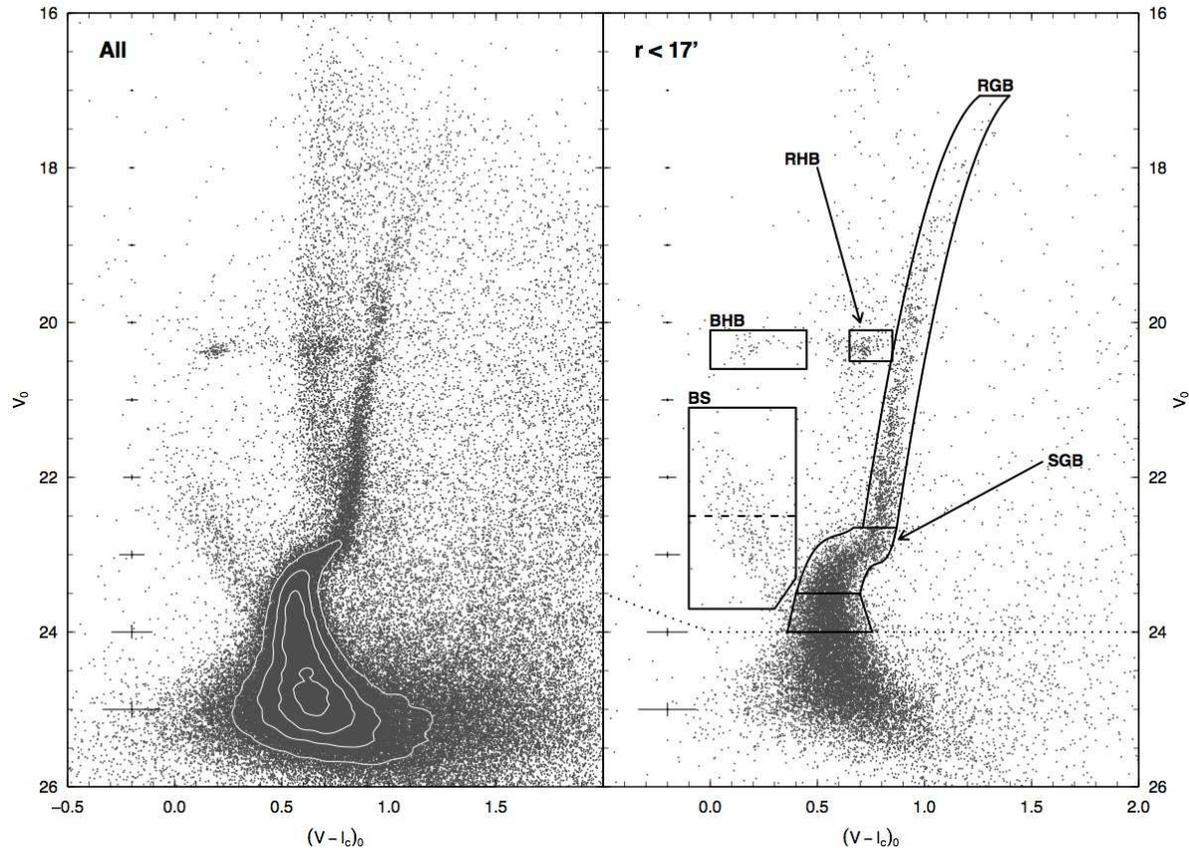}
 \caption{The colour-magnitude diagram of the central $80\arcmin \times 100\arcmin$ and within r=17\arcmin area of Sextans dSph.  Error bars show the photometric error at each magnitude level based on the artificial-star test. The boundaries marked in the left panel are used to select as sub-giant branch (SGB), red giant branch (RGB), blue and red horizontal branch (BHB, RHB) and blue straggler (BS) stars belonging to Sextans.}
  \label{fig:3}
\end{figure*}

We checked the consistency of the magnitude of stars commonly detected in both long and short exposure images (typically 20-22 mag stars), and in both side by side images.  The magnitude difference between the long and short exposures is $\Delta M_{short-long} < 0.02$, and between the side by side images is $\Delta M_{neighbour} < 0.05$.  The Galactic extinction is taken from \citet*{1998ApJ...500..525S} for the direction of each observed field in Sextans.  The assumed extinction law is the R$_v$=3.1 \citep{1989ApJ...345..245C} and the relation of A$_I$/A$_V$=0.594 \citep{1998ApJ...500..525S}.

The completeness and photometric errors were derived using the artificial-star test with the ADDSTAR routine in DAOPHOT.  We added 7000 artificial stars to each image in every 0.5 magnitude interval from 18 mag to 24 mag and in every 0.25 magnitude interval from 24 mag to 26 mag.  We processed the resulting images containing artificial stars in the same way as for the original images to estimate the detection ratios.  The ratio, N(recovered)/N(added) of the best seeing field (R05) and the worst seeing field (R25) are plotted in Figure \ref{fig:2}, which indicates that our photometry is at least 90$\%$ complete at 24 mag in both bands covering the entire region.  The mean photometric errors plotted in Figure \ref{fig:3} in Section 3 are based on the difference between the input magnitude and the output magnitude of the artificial stars.

\section{Colour-magnitude diagram and structural properties}

\subsection{Colour-magnitude diagram}

Figure \ref{fig:3} presents the de-reddened CMD of the star-like objects found in the central nine fields (R01, R02, R07, R12, R18, R24, R25, R26, R27; hereafter the core region) that cover the central $80\arcmin \times 100\arcmin$ area of the Sextans dSph.  Error bars show the photometric errors at each magnitude with $\vio=0.75$ based on the artificial star test.  
The CMD contains approximately 74,000 sources.  It shows a well-populated RGB extending downward from $\vo\sim17$, the red and blue HB (RHB, BHB) at $\vo=20.3$, the well-defined sub-giant branch (SGB) to MS traced below $\vo\sim 23$, together with the numerous BS candidates.  All the features show that the Sextans is basically dominated by old metal-poor population, but the existence of both RHB and BHB and the thickness of SGB and MS suggest that there are multiple populations in this galaxy (see Section 4).  Unfortunately, stars brighter than $I_c \sim 16$ were saturated in our $I_c$-band images, so that the tip of RGB ($I_{c,TRGB}=15.95\pm0.04$ in L03) can hardly be identified.  Figure \ref{fig:3} includes foreground Galactic stars which mainly distribute uniformly at $\vo>0.6$, and background unresolved objects.  To estimate the significance of the contamination, we use A01 and A03 fields which located at the outside of the tidal radius of \cite{1995MNRAS.277.1354I}. 

\begin{figure}
 \includegraphics[width=250pt,clip]{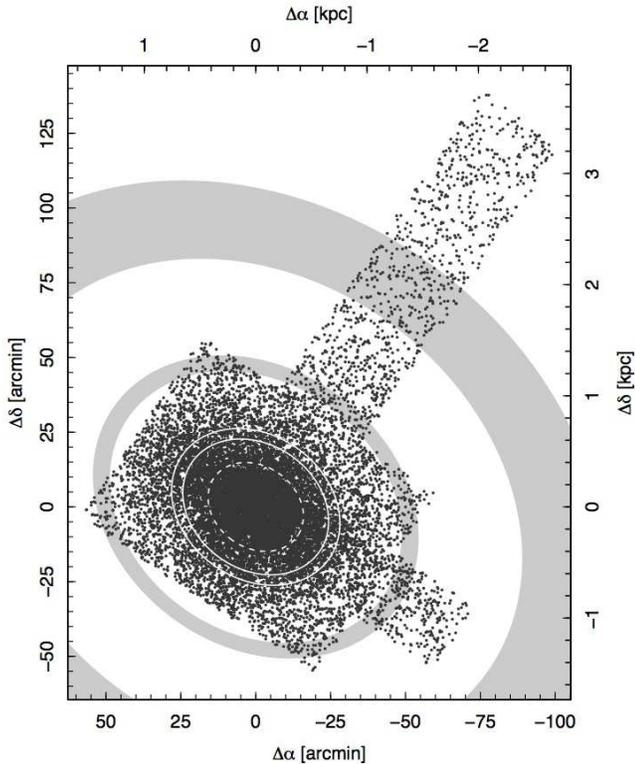}
 \caption{The spatial distribution of the member stars of Sextans.  The plotted stars are located within the boundaries in the CMD shown in Figure \ref{fig:3}.  The areas within the central dashed white line ($\rm{r}=\rc$), between two solid white lines ($\rm{r} \sim \rh$), and within the gray coloured regions ($\rm{r} \sim 2\times\rh$, $\rm{r} \sim \rt$) are used in the detailed analysis of the stellar components at various distances in Figure \ref{fig:6}.}
  \label{fig:4}
\end{figure}

\subsection{Structural properties}
To investigate the distance to Sextans, we use the $V$-band magnitude of the HB stars.  We estimate the HB magnitude as $V_{0, HB} = 20.33 \pm 0.02$ derived from the 391 RHB stars with $20.1 < \vo < 20.5$ and $0.55 < \vio < 0.85$, and the 139 BHB stars with $20.1 < \vo < 20.5$ and $0.0 < \vio < 0.45$ in the core region.  The metallicity distribution of the Sextans dSph was determined by \cite{2011MNRAS.411.1013B}.  We adopt $\feh=-2.0$ for the average to assume a theoretical value of the absolute HB magnitude to be $M_{V,\rm{HB}}=0.496$. We then obtain a distance modulus $\mm = 19.83 \pm 0.05$ (corresponding to a distance of $92.5 \pm 2.2$ kpc).  This value agrees with the previous estimates of L03; they gave $\mm = 19.90 \pm 0.06$ from the $I_c$-band magnitude of TRGB, and $\mm = 19.89 \pm 0.04 $ based on the HB magnitude.  

Figure \ref{fig:4} shows the spatial distribution of the star-like objects within the boundaries in Figure \ref{fig:3}, which were drawn by hand. The stars within the areas marked by white lines and gray colour are used to investigate the CMD morphologies at various distances in Section 4. In order to estimate structural properties, we chose these member stars in the core region ($80\arcmin \times 100\arcmin$, equivalent to $2.1 \times 2.4$ kpc), to reduce contamination and to avoid the incompleteness of the photometry and the difference of completeness between each region.  The centroid, ellipticity and position angle are derived from the density-weighted first and second moments and are listed in Table \ref{tbl:2}.    

\begin{figure}
 \includegraphics[width=170pt, angle=-90]{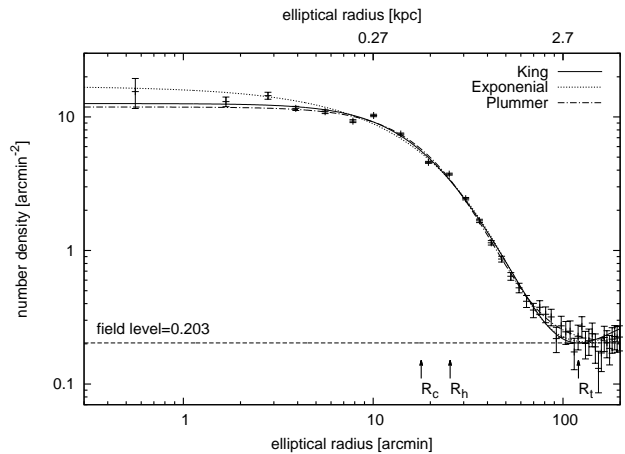}
 \caption{Radial profile derived by calculating the average number density within elliptical annuli.  Best-fitting King profile with core radius $\rc = 17.91\arcmin \pm 0.65\arcmin$ and tidal radius $\rt = 120.5\arcmin \pm 7.7\arcmin$, exponential profile with the effective radius $r_{e} = 15.15\arcmin \pm 0.30\arcmin$ and Plummer profile with the Plummer radius $b = 26.60\arcmin \pm 0.43\arcmin$ are overlaid as solid, dashed and dotted line, respectively. }
  \label{fig:5}
\end{figure}

\begin{table}
\caption{The structural properties of Sextans dSph}
\label{tbl:2}
\begin{tabular}{@{}lc}
\hline\hline
	Parameter & Value \\
\hline
	Coordinates (J2000) & 10h13m04.9s, $-01^{\circ}37\arcmin31.6\arcsec$\\
	Position angle of major axis  & $54^{\circ}.2$\\
	Ellipticity & $0.20$\\
	$\mm$ & $19.83 \pm 0.05$\\
	Distance & $92.5 \pm 2.5$ [kpc]\\
	$\rc$ & $481 \pm 17$ [pc]\\
	$\rt$ & $3.24 \pm 0.2$ [kpc]\\
	$\rh$ (exponential) & $685 \pm 14$ [pc]\\
	$\rh$ (plummer) & $715 \pm 12$ [pc]\\
\hline
\end{tabular}
\end{table} 

In Figure \ref{fig:5}, the stellar radial profile in the logarithmic form is constructed by calculating the average number density of member stars in elliptical annuli with the ellipticity and the position angle derived above.  The field contamination level is estimated from the A01 field.  The observed area within each elliptical annulus is calculated by the Monte Carlo method.  The error bars take into account the Poisson errors, the uncertainties in the field level estimation, and the area estimation.  We fit the radial profile with standard King, exponential and Plummer models using a least-squares minimization technique.  The best-fitting models are overlaid on the radial profile in Figure \ref{fig:5} as solid, dashed and dotted lines, respectively.  The derived parameters, the core radius $\rc$, the tidal radius $\rt$ and the half-right radius $\rh$ of the exponential and Plummer profiles are listed in Table \ref{tbl:2} with the standard error of the model fitting.

\begin{figure*}
 \includegraphics[width=450pt,clip]{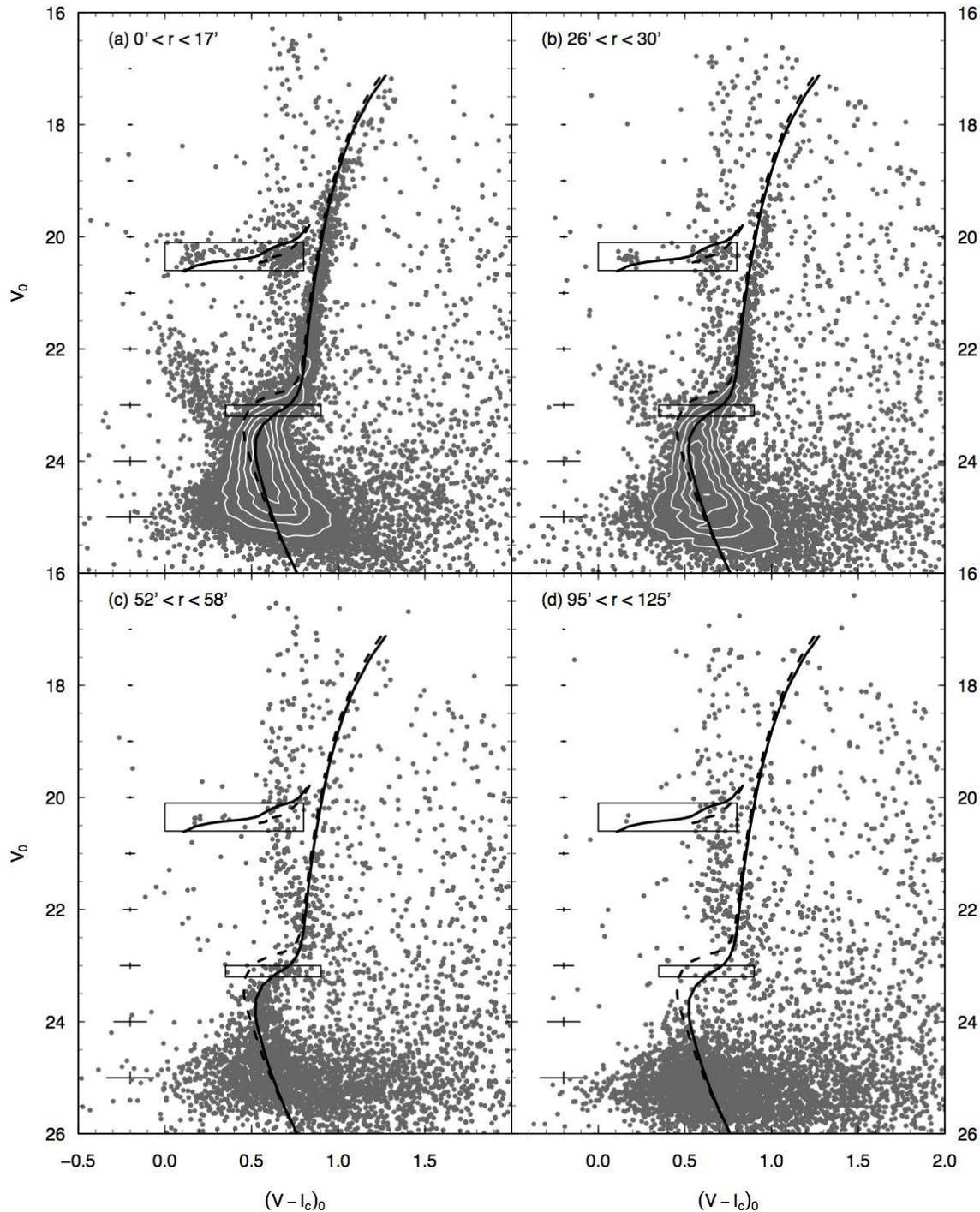}
 \caption{The CMDs of stars at various distances from the centre of Sextans.  The theoretical isochrones of 10.0 Gyr and 13.8 Gyr with Z=0.0002 are overlaid as dashed and solid lines, respectively.  The distance from the centre of each CMD is described in each panel. The stars within the solid rectangles are used to derive the colour distribution of HB and SGB stars in Figure \ref{fig:7} and \ref{fig:8}.}
  \label{fig:6}
\end{figure*}

Our estimation of the structural properties agree with \cite{1995MNRAS.277.1354I} and \cite{2016MNRAS.460...30R}, except for the ellipticity and the tidal radius.  Our ellipticity $e=0.20$ is slightly smaller than their estimations.  Although it is within the margin of error. The estimated tidal radius $\rt=120.5\arcmin \pm 7.7\arcmin$ is smaller than \cite{1995MNRAS.277.1354I} estimation of $\rt=160\arcmin \pm \arcmin 50$, and larger than \cite{2016MNRAS.460...30R}'s $\rt=83.2\arcmin \pm 7.1\arcmin$.  These differences are mainly due to the area coverage and the image depth.  They covered the entire region of Sextans, while we use deeper images of the central area.  The best-fitting King model in the left panel of Figure 8 of \cite{2016MNRAS.460...30R} does not fit the data at large radii well, implying a larger tidal radius. The contour map of \cite{1995MNRAS.277.1354I} shows that the inner parts of Sextans are less elliptical than the outer parts.  Our tidal radius estimation is based on the images extended toward the minor axis.  Therefore, if we assume that the outer ellipticity is $e=0.35$, the tidal radius increases to $\rt=148.3\arcmin \pm 9.4\arcmin$, which agrees well with \cite{1995MNRAS.277.1354I}.

\section{The stellar populations}

\begin{figure}
 \includegraphics[width=250pt,clip]{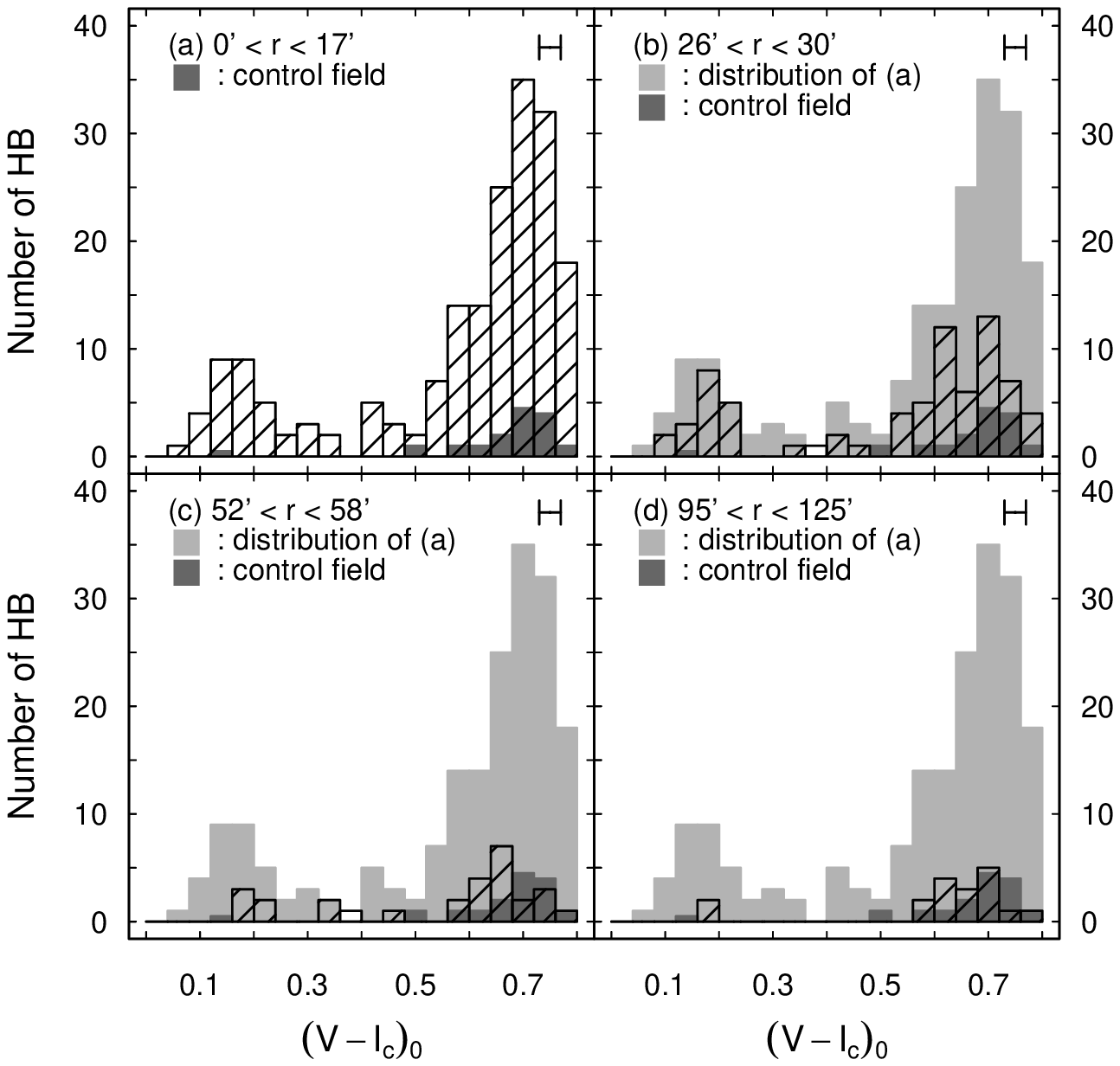}
 \caption{The colour distribution of the HB stars within $20.1<\vo<20.6$ in each CMD of Figure \ref{fig:6} is shown as the shaded bar.  The width of colour bin is 0.04 mag.  The histograms with light gray colour in \ref{fig:7}b-\ref{fig:7}d indicate the distribution of \ref{fig:7}a as the basis.  The dark gray histogram shows the distribution of the field contamination estimated by using the region of $135\arcmin < \rm{r} < 175\arcmin$ from the centre of Sextans dSph.  The error bar shows the photometric error, $(V-I_c)_{0,err}=0.02$, at $\vo=20.5$ and $\vio=0.5$.}
  \label{fig:7}
\end{figure}

\begin{figure}
 \includegraphics[width=250pt,clip]{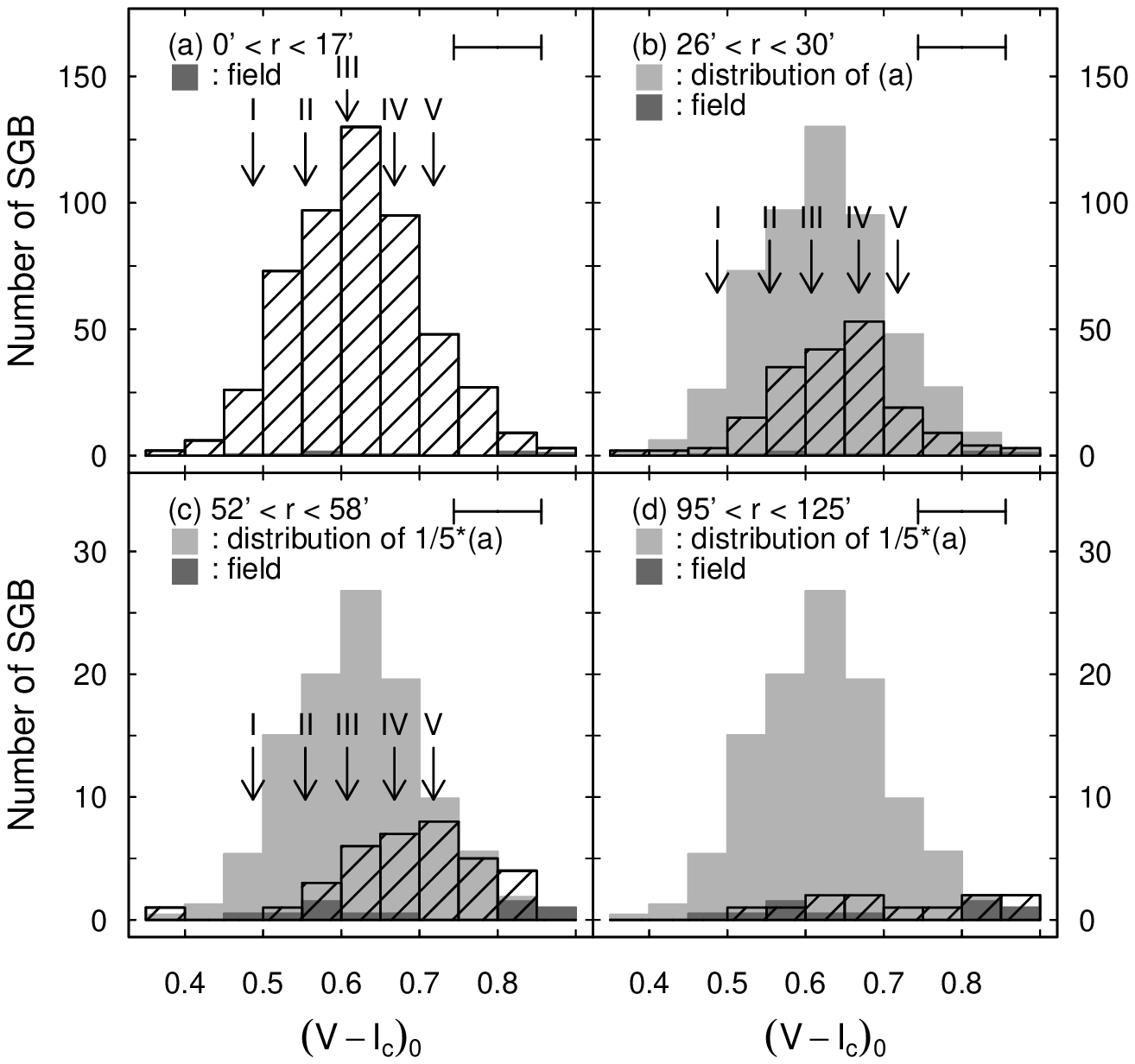}
 \caption{The colour distributions of the faint SGB stars within $23.0<\vo<23.2$ in the CMDs of Figure \ref{fig:6} are shown as the same manner as Figure \ref{fig:7}.  The histograms with light gray colour in Figure \ref{fig:8}c and \ref{fig:8}c show the distribution of Figure \ref{fig:8}a with a scale factor of 1/5.  The error bar shows the average photometric error, $(V-I_c)_{0,err}=0.05$, at $\vo=23.0$ and $\vio=0.5$. The arrows I to V indicate the theoretical colours of 10.0, 11.2, 12.6, 13.8, and 15.5 Gyrs old stars, respectively.}
  \label{fig:8}
\end{figure}

Stellar populations of the ``classical" dSphs are more complicated than fainter dwarf galaxies.  Bright dSphs are known to possess multiple and spatially distinct stellar components in the RGB and HB \citep[e.g.][]{2004ApJ...617L.119T, 2006A&A...459..423B, 2008ApJ...681L..13B}.  Fornax and Sculptor dSphs ($\mv=-13.4,-11.1$) are confirmed to have the population gradients; the old, metal-poor populations exist at all radii, while more metal-rich, younger stars are more centrally concentrated \citep{2012A&A...544A..73D, 2012A&A...539A.103D}.  Carina dSph shows at least three distinct MS turn-offs \citep{2003AJ....126..218M}, and the faintest ``classical" dSph, Sextans also shows the different spatial distribution of RHB and BHB stars (L03).  

To demonstrate the spatially difference of the stellar population in Sextans, we compare the CMDs of star-like objects in the elliptical annuli at various distances.  Figure \ref{fig:6} shows the CMDs of the four regions located at the elliptical distance of (a) $0 < \rm{r}< \rc$, (b) $\rm{r} \sim \rh$, (c) $\rm{r} \sim 2\times\rh$, (d) $\rm{r} \sim \rt$, respectively.   The selected regions cover almost the same solid angle (530 to 550 $\rm{arcmin}^2$), hence, the number of foreground/background objects in four CMDs are considered to be similar.  Padova isochrones of 10.0 Gyr and 13.8 Gyr with Z=0.0002 are overlaid as guides \citep{2008A&A...482..883M}.  

\subsection{The spatial difference of CMD morphology}

In the innermost region, Figure \ref{fig:6}a shows the dominant RHB, a few BHB, the relatively wide sequence of SGB to MSTO, and numerous BS stars.  The overlaid 10 Gyr and 13.8 Gyr isochrones well reproduce the distribution of stars.  At the half-light radius (Figure \ref{fig:6}b), the clump of RHB fades into the foreground contamination, and the width of MSTO and SGB becomes narrow.  The younger isochrone could not fit to MSTO to SGB which seems to be shifted to redder and fainter than that of the inner region.  The differences between Figure \ref{fig:6}a and Figure \ref{fig:6}b still remain when we re-sample the data as the same size. Figure \ref{fig:6}c is similar to Figure \ref{fig:6}b, but there are only a few stars.  The RHB, RGB, BS stars almost disappear, but the BHB and the MS stars can be identified.  In Figure \ref{fig:6}b and Figure \ref{fig:6}c, even the oldest (13.8 Gyr) isochrone shows a slightly bluer colour at MSTO than that of Sextans dSph.  If we adopt the more metal-rich isochrone (Z=0.0004) for these outer regions, this difference can be reduced, but it is not likely that the spectroscopically confirmed metallicity is biased significantly \citep{2011MNRAS.411.1013B}.  Figure \ref{fig:6}d includes the star-like objects at the edge of Sextans dSph ($\rt=124.7\arcmin \pm 7.8\arcmin$).  Most of the components can hardly be identified. 

In these four CMDs, the HB and SGB colour distributions change with the distance from the galaxy centre.  Figure \ref{fig:7} and Figure \ref{fig:8} show the colour distributions of HB and SGB stars within the rectangles in Figure \ref{fig:6}, which are selected to enhance color differences with respect to age of the populations and to minimize field contamination. As a guide, the histograms with light gray colour in Figure \ref{fig:7}b-d and in Figure \ref{fig:8}b are the same as that of the innermost region, and those in Figure \ref{fig:8}c and \ref{fig:8}d are the distribution of the innermost region with a scale factor of 1/5.  

While the BHB stars are seen in all panels in Figure \ref{fig:7}, the RHB stars dominate within $\rc$, and decrease rapidly with the distance.  This spatial difference of HB distribution implies that the old/metal-poor population exist in the whole region of Sextans dSph, while the relatively younger/metal-rich population is situated only in the central region as shown in L03.

On the other hand, the peak of the $\vio$ colour of SGB distribution becomes redder from the centre ($0.62$ in $\rm{r}<\rc$) to the outer region (0.70 at $\rm{r}=\rh$, 0.72 at $\rm{r}=2\times\rh$), and the bluer part of the SGB disappears in Figure \ref{fig:8}c and Figure \ref{fig:8}d. This difference is larger than the photometric error at this magnitude. The bluer SGB stars in the inner region indicate either a relatively younger or more metal poor population compared to the outer region. However, since spectroscopic studies show a metallicity gradient from the metal-rich central part to the metal-poor outer region in Sextans dSph, the latter case is unlikely \citep{2011MNRAS.411.1013B, 2011ApJ...727...78K}.  The theoretical colours of 10.0, 11.2, 12.6, 13.8, and 15.5 Gyrs old stars with $\feh = -2.0$ (Z=0.0002 with $\afe=0.0$) at the magnitude based on Padova isochrones are shown as arrows I to V in Figure \ref{fig:8}.  The peaks between the innermost ($\rm{r}<\rc$) and the outer ($\rm{r}\sim 2\times\rh$) region corresponds to about $\sim$ 3 Gyr age difference, assuming the constant metallicity of $\feh =-2.0$.  

Note that if we adopt a metallicity gradient as found by previous spectroscopic observations, the spatial age difference estimated here would become larger. This is due to the fact that for a given age, the color of SGB stars increases with metallicity, therefore a metal rich population needs to have a much younger stellar age in order to produce a blue color distribution. 
%Since more metal-rich SGB stars show redder color at the same age, if there are more metal-rich stars in the inner part, these stars should have an even younger age than the case of the constant metallicity for having this bluer color distribution. 

\subsection{The population gradient}

The spatial difference of the CMD morphology reveals that the stellar population of Sextans is different from inside to outside out to $2\times\rh$.  The innermost part possess both young/metal-rich and old/metal-poor populations which appear in the HB morphology and color distribution of the SGB.  While younger component fade away, the older population still exists beyond the half light radius.  Although the age and metallicity of stars are degenerate in colour and magnitude of HB and SGB to MSTO, the spatial differences of both HB and faint SGB colour distributions in the previous subsection indicate the existence of the radial age gradient on the assumption of the constant metallicity or the metallicity gradient in the Sextans.  This spatial difference could also be seen in the radial profile of the RHB, BHB, and SGB stars.  

\begin{figure}
 \includegraphics[width=240pt]{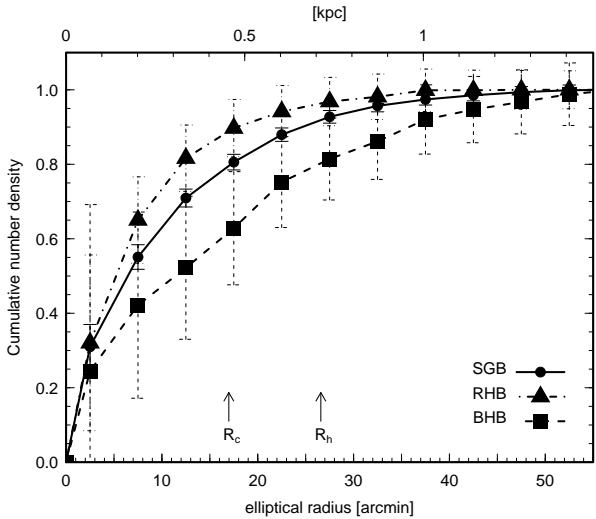}
 \caption{Cumulative radial profile of stellar components derived by calculating the average number density within elliptical annuli.  The triangles, squares, and circles represent the profiles of RHB, BHB and SGB stars, respectively. The overall core and half-light radii are also indicated.}
  \label{fig:9}
\end{figure}

Figure \ref{fig:9} shows the cumulative radial number density profiles.  Each star is selected by the colour and magnitude criteria in Figure \ref{fig:3}.  In Figure \ref{fig:9}, the distribution of BHB stars is more extended than SGB component, and the RHB is concentrated toward the center.  The SGB profile is inbetween BHB and RHB, because SGB includes both centrally concentrated young stars and spatially extended old populations.  The RGB and BS stars have similar profiles to SGB stars.  

\begin{figure}
 \includegraphics[width=240pt,clip]{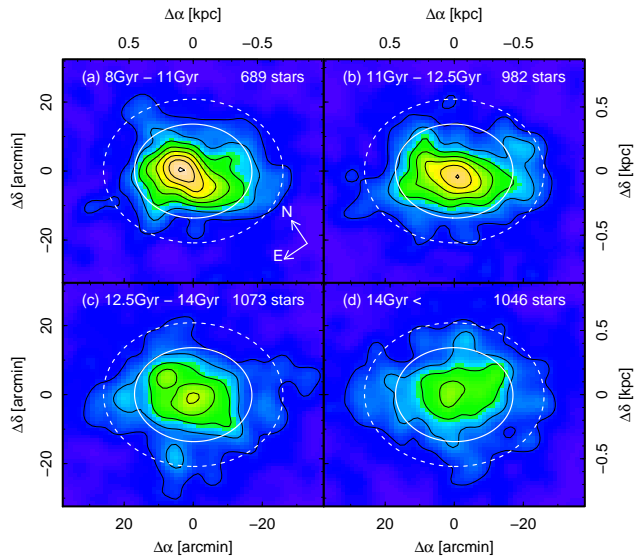}
 \caption{The contour maps of SGB stars in the different age bins. The density levels and colour scales of panels (b)-(d) are fixed to that of panel (a). The solid and dashed white lines show the core and half-light radius, respectively. }
  \label{fig:10}
\end{figure}

Spectroscopic observations show a wide range of metallicity dispersions and a metallicity gradient in Sextans dSph \citep{2011MNRAS.411.1013B, 2011ApJ...727...78K}.  However, if we assume the constant metallicity as $\feh = -2.0$, the spatial distribution of stars within a specific age range could be investigated using theoretical isochrones.  Figure \ref{fig:10} shows the spatial distributions of SGB stars within the magnitude range $22.7 < \vo < 23.5$, which is bright enough to ignore the spatial disparities of the photometric errors and the completeness, and is adjusted to reduce the foreground/background contaminations.  We divide SGB stars into the age bins of (a) 8 to 11 Gyr, (b) 11 to 12.5 Gyr, (c) 12.5 to 14 Gyr, and (d) 14 to 15.5 Gyr using Padova isochrones.  Then these stars are binned and smoothed by the Gaussian kernel with the bandwidth of 2\arcmin5 to draw the spatial contour maps in Figure \ref{fig:10}.  The density scales in panel (b)-(d) are fixed to that of the panel (a).  The foreground/background contaminants are almost negligible ($< 3\%$).  

In Figure \ref{fig:10}, the younger SGB population has higher concentration and higher ellipticity than the old population.  The density peak in the panel (a) has slight offset from the galaxy centre.  There is a secondary peak at the northeast side of $\rc$ in the panel (c) that has similar position to the cold substructure found by \cite{2006ApJ...642L..41W}.  We cannot confirm this secondary peak in the density map of all member stars in Figure \ref{fig:4}, and there is no clear difference of the CMD morphology between this area and the surrounding area. But if it is the remnant of a disrupted cluster, the stellar age is as old as UFDs and old Galactic globular clusters.  

The radial population gradients and two distinct dynamical components are found in many bright dwarf satellites.  The brightest UFD, Canes Venatici I (CVn I) also contains radially different stellar populations; the distribution of RHB stars is more concentrated toward the centre than that of BHB stars \citep{2012ApJ...744...96O}.  On the other hands, faint satellites, such as Bo\"otes I, have no spatial difference in the CMD morphology, and are all consistent with an age of ($\sim 13$Gyr). From faint UFDs to bright dSphs, the complexity of stellar populations increase in proportion to the increasing total luminosity.  The stellar population structure of Sextans dSph follows this tendency.  Sextans dose not show obvious evidence of young or intermediate age population as the Sagittarius and Fornax dSphs, but shows the age range larger than 3 Gyr at the inner most region. 

Principally, stars in satellite galaxies were born in very low mass halos.  All the Galactic dSphs contain the oldest population comparable to those of the old Galactic globular clusters, which suggests that the local over-densities in the primordial matter distribution started forming stars at a similar time.  Then, the gaseous matter in a lower mass halo was easily blown out by the supernova feedback when the star formation occurred, or was possibly photo-evaporated by reionization \citep[e.g.][]{2009MNRAS.400.1593M, 2011MNRAS.413..101G}.  

After UFDs stopped their star formations, some halos kept the gas and continued the star formation for a few Gyrs then became brighter systems.  \cite{2012MNRAS.425..231R} have 
calculated the probability distribution of the infall time for the Milky Way satellites based on the cosmological simulation.  According to their estimates, Sextans was accreted between 7 and 9 Gyr ago, which was a similar time as the end of forming stars at the inner most region.  It implies that the star formation in Sextans was gradually quenched from outside to inside, then finally stopped when it fell into the Milky Way.  Or, multiple pericentric passages around the Milky Way could removed gas gradually from Sextans, which also made the population gradient \citep{2014A&A...564A.112N}.  

\section{Blue straggler stars}
Blue straggler (BS) candidates are found universally in the Galactic dSph satellites.  The dSphs have predominately old populations, but they might have had a low-level star formation in last few Gyrs at the inner region.  Therefore, unlike BS stars in the old globular clusters (GC), these stars can be either genuine BS stars or substantially young MS stars ($\sim 2$Gyr).  L03 investigated the BS stars within central $20\arcmin$ of Sextans and showed that the bright BS stars are more centrally concentrated than the faint BS stars.  

In this section, we probe the properties of BS stars populated throughout the galaxy.  The criterion for BS stars is shown in Figure \ref{fig:3}.  The magnitude and colour ranges are similar to those of L03, but we modify the boundary of the minimum magnitude.   Thanks to the deeper photometry, we can easily discriminate BS stars against MS stars and the contaminations at this magnitude level.  

\subsection{The radial distribution}
In our catalogue, there are 288 BS stars within the core radius, and 747 BS stars in the observed region within the tidal radius.  The contamination level is estimated using the control field, about $2\%$ within the core radius and $10\%$ within the tidal radius, respectively.  
We note that there are two BS stars within $\rm{r}=100\arcsec$ and within the selection criterion of L03.  These stars have not been found in L03, because they are located at the CCD gap of their images. 

\begin{figure}
 \includegraphics[width=240pt,clip]{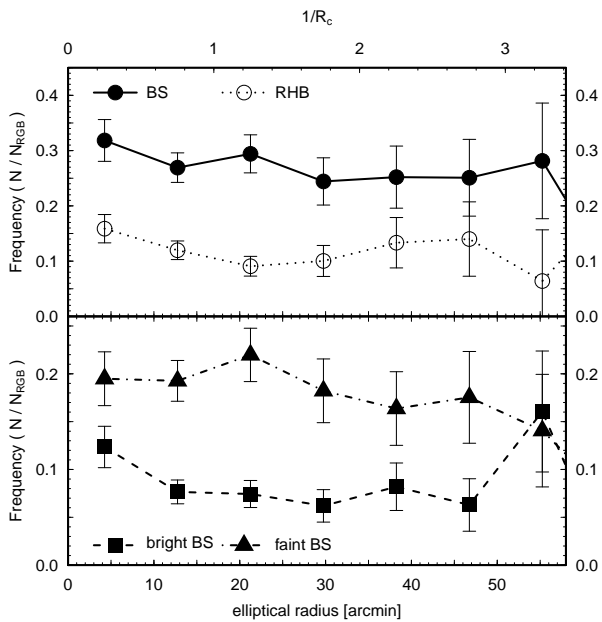}
 \caption{The radial distributions of the relative frequencies of BS stars (filled circles connected by solid line), of RHB stars (open circles connected by dotted line), of bright BS stars (filled squares connected by dashed line) and of faint BS stars (filled triangles connected by dotted-dashed line).  The numbers of stars within elliptical annuli are counted and corrected for the contamination and normalized to those of RGB stars. }
  \label{fig:12}
\end{figure}

To investigate the radial difference between BS and other populations, and between bright ($\vo<22.5$) and faint ($\vo>22.5$) BS stars, we plot the relative frequencies of BS stars normalized to RGB stars in Figure \ref{fig:12}, together with that of RHB stars to RGB stars. The numbers of stars within elliptical annuli are counted and corrected for the contamination, using the reference field outside of the tidal radius. The errors include Poissonian statistics, uncertainties in foreground subtraction and photometric errors.  

In Figure \ref{fig:12}, the radial frequency of $\rm{N}_{\rm{BS}}$ with respect to $\rm{N}_{\rm{RGB}}$ is nearly flat and similar to that of $\rm{N}_{\rm{RHB}}$ to $\rm{N}_{\rm{RGB}}$, showing the spatial distribution of the BS stars is similar to those of the RGB and RHB stars. If the considerably younger MS stars ($\sim 2$Gyr) exist in Sextans and are found as BS candidates, they would be expected to be centrally located, rather than distributed throughout the galaxy.  Considering with the long dynamical timescale and the low stellar density, the uniform BS distribution indicates that they are the mass-transfer BS stars evolved from primordial binaries, as those found in other dSphs \citep{2007A&A...468..973M, 2007MNRAS.380.1127M,2012ApJ...744...96O}.  

As pointed out by L03, the relative frequency of the bright BS stars (filled squares) seems to be decreasing slightly with increasing radius up to at $\rm{r} \sim \rc$, while that of the fainter BS stars (filled triangles) shows the flat distribution.  Note that we compare the radial frequency with that of BS stars based on the L03 definition and do not see any significant difference.

\subsection{The luminosity distribution}

\begin{figure}
 \includegraphics[width=240pt,clip]{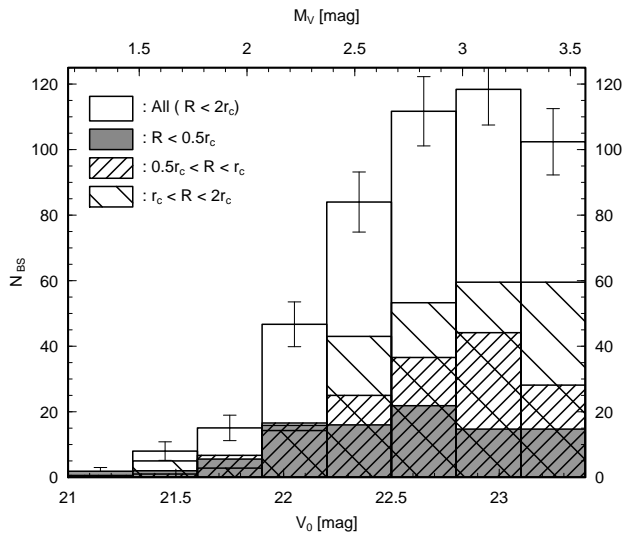}
 \caption{The luminosity distribution of BS stars. The empty histogram represents the all sample of BS stars within $\rm{r}<2\times\rc$, and the error bars show the Poissonian errors. The gray coloured, heavily and lightly hatched histograms represent BS stars located at $0<\rm{r}<0.5\times\rc$, $0.5\times\rc < \rm{r} <\rc$, and $\rc < \rm{r} < 2\times\rc$, respectively.}
  \label{fig:13}
\end{figure}

To probe the radial gradients of the bright BS frequency in the inner region, we compare the luminosity distribution of BS stars in the innermost region ($\rm{r} < 0.5\times\rc$) with those of BS stars located in the outer regions ($0.5\times\rc < \rm{r} < \rc$ and $\rc < \rm{r} < 2\times\rc$).  The foreground and background contaminations are corrected using the field outside of the tidal radius.  In Figure \ref{fig:13}, the number of the central BS stars (gray coloured histogram) increases rapidly with decreasing luminosity and becomes constant at the magnitudes fainter than $\vo=22$, while those of BS stars in the outer regions (heavily and lightly hatched histograms) increase gradually and keep increasing at fainter magnitudes.  At the magnitude $\vo=22$, one third of BS stars are located within $\rm{r}<0.5\times\rc$.  As the luminosity decreases, the proportion of central BS stars to all BS stars decreases and it becomes $12\%$ at $\vo=23$.  The different luminosity distributions are confirmed by the two-sample Kolmogorov-Smirnov test.  A similar correlation between the radial distance and the luminosity distribution was also found in Fornax dSph, but not found in Draco, Ursa Minor and Sculptor \citep{2009MNRAS.396.1771M}. 

The luminosity distribution of BS stars located at the outside of the core radius (lightly hatched histogram) is not significantly different from that of BS stars at $0.5\times\rc < \rm{r} < \rc$ (heavily hatched histogram) in Figure \ref{fig:13}.  We note that the luminosity distribution of BS stars beyond $2\times\rc$ has also similar distribution that keeps increasing with decreasing luminosity. 

Within the innermost region ($\rm{r}< 0.5\times\rc$), we confirm that the brighter BS stars are more centrally concentrated than the fainter ones, which can be the result of either the mass segregation, the existence of younger population at the central region, or the remnant of a disrupted stellar cluster. However, the first one is doubtful, because of the long dynamical relaxation timescale of Sextans dSph.  

A possible explanation is that the inner brighter BS stars are younger than the fainter ones. Typical BS stars formed via mass transfer are massive than the MSTO mass, but not more than twice the MSTO mass \citep[e.g.][]{1964MNRAS.128..147M,1997ApJ...489L..59S}.  
Although it is difficult to determine the mass of bright BS stars, the centrally concentrated younger population having higher MSTO mass would contain the brighter BS stars than those of the old population.

\cite{2004MNRAS.354L..66K} found a kinematically cold substructure in the central $5\arcmin$, which may be a remnant of a disrupted star cluster which sank into the Sextans centre.  It was confirmed with a larger spectroscopic sample and its total luminosity was estimated as $2.2\times 10^{4}\lsolar$ \citep{2011MNRAS.411.1013B}.  In a high density region, such as a star cluster, bright BS stars are thought to be created preferentially through the direct collision of two stars, while it would not be happen in a low density system \citep[e.g.][]{1995ApJ...439..705B}.  Therefore, 
a part of centrally concentrated bright BS stars could be originally formed in a star cluster that was disrupted and sunk into the Sextans centre.  By subtracting the primordial BS luminosity distribution estimated from the outer region, the number of BS stars of cluster-origin, in $\rm{r}< 0.5\times\rc$, can be estimated as $N= 17.9\pm1.2$, which is the typical BS number in a star cluster of $10^{4}\lsolar$ \citep{2013ApJ...774..106S}.  

Although, only spectroscopic analyses can provide the indisputable evidence to reveal the origin of bright BS distribution in Sextans, our results are consistent with both scenarios.

\section{Conclusions}

We present the wide-field, deep photometry of the Sextans dSph, sampling fields extended to the tidal radius and reaching two magnitude below the main-sequence turn-off.  It enables us to derive the global properties of stellar populations.  We estimate the distance, structural properties and probe the differences of the stellar populations with the distance from the galaxy centre. 

The analysis of the radial distribution of each evolutionary phase shows that blue HB stars have more spatially extended distribution, while red HB stars are more centrally concentrated.  The colour distributions of SGB stars also show the spatial difference; the SGB stars in central regions present bluer colour than those in the outskirt.  

These results indicate the age gradient of Sextans. The relatively younger stars ($\sim10$ Gyr) are more centrally concentrated than the old stars ($\sim13$ Gyr), and the star formation in the innermost region continued more than $\sim3$ Gyr.  The spatial maps of SGB stars in different age bins also confirm that the younger population has the higher concentration and higher ellipticity than the old population.  Considering the long dynamical relaxation timescale, these features may have kept effects of the self-regulation through stellar feedback during the star formation epoch. 

The BS stars in the Sextans show the similar radial distribution to those of RGB and RHB stars, indicating that they are the genuine mass-transfer BS stars evolved from primordial binaries, as those found in other dSphs.  In the innermost region ($\rm{r} < 0.5\times\rc$), the number of BS stars increases rapidly with decreasing luminosity and becomes constant at fainter magnitude, while those of BS stars in other regions increase gradually.  This centrally concentrated bright BS distribution can be interpreted as the results of either the existence of younger population at the central region or the collisional BS stars belonging to the remnant of a disrupted star cluster of $10^{4}\lsolar$ sank into the Sextans centre.  

Sextans has the large tidal radius and a low surface brightness, so it could be the best target to probe the evidence of tidal disruption by exploring substructures out to the tidal radius. \cite{2016MNRAS.460...30R} investigated the entire area of the Sextans and found no sign of tidal disruption.  We also see no tidal features such as S-shaped contours and tails in our footprint.  The existence of the age gradient suggests that it is unlikely that the inner parts of Sextans have been perturbed by any strong tidal disruption. 
Further observations, especially multi-object spectroscopy carried by super wide-field equipments on large telescopes will help us to demonstrate all the details of the outskirts of Sextans and trace their chemical and dynamical properties.

\section*{Acknowledgments}
We are grateful to the entire staff at Subaru Telescope.  We acknowledge the importance of Maunakea within the indigenous Hawaiian community and with all respect say mahalo for the use of this sacred site.  SO acknowledges support from the CAS PIFI scheme.

\bibliographystyle{mn2e}

\label{lastpage}

\end{document}